\documentclass[preprint,preprintnumbers,nobibnotes,pra,showkeys,onecolumn,secnumarabic,showpacs]{revtex4}
\usepackage{amsmath}
\usepackage{graphicx}
\usepackage{dcolumn}
\usepackage{bm}

\input{tcilatex}
\begin{document}

\title{Effects of random fields on the dynamics of the one-dimensional
quantum XXZ model}
\author{Yin-Yang Shen$^{a}$ Xiao-Juan Yuan$^{a,b}$}
\author{Xiang-Mu Kong$^{a}$}
\altaffiliation{Corresponding author.\\
E-mail address: kongxm@mail.qfnu.edu.cn (X.-M. Kong).}
\affiliation{$^{a}$Shandong Provincial Key Laboratory of Laser Polarization and
Information Technology, Department of Physics, Qufu Normal University, Qufu
273165, China\\
$^{b}$School of Physics, Shandong University, Shandong 250100, China}

\begin{abstract}
The dynamics of the one-dimensional spin-1/2 quantum XXZ model with random
fields is investigated by the recurrence relations method. When the fields
satisfy the bimodal distribution, the system shows a crossover between a
collective-mode behavior and a central-peak one with increasing field while
the anisotropy parameter $\Delta $ is small; a disordered behavior replaces
the crossover as $\Delta $ increases. For the cases of Gaussian and
double-Gaussian distributions, when the standard deviation is small, the
results are similar to that of the bimodal distribution; while the standard
deviation is large enough, the system only shows a disordered behavior
regardless of $\Delta $.
\end{abstract}

\maketitle

\section{INTRODUCTION}

Quantum spin systems are of considerable interest for the reasons that they
provide a ground for studying quantum many-particle phenomena, and offer the
possibility to compare the theoretical and experimental results, etc. The
dynamical properties of the quantum spin systems have received much
attention and a variety of achievements have been got over the years. Among
the systems studied, the one-dimensional (1-D) spin-1/2 XXZ chain is one of
the nontrivial models and has been used to describe several quasi-1-D
compounds such as CsCoBr$_{\text{3}}$ \cite{3,3aa,5}, CsCoCl$_{\text{3}}$ 
\cite{4,5}, Cs$_{\text{2}}$CoCl$_{\text{4}}$ \cite{1,2} and TiCoCl$_{\text{3}%
}$ \cite{6}. Concerning the dynamics of this model, the spin correlation
function has been studied by using approximation and numerical methods \cite%
{A,B,C,D,E,F,G,Bohm,Lee,Lee2}. Strong numerical evidence was found for a
change of the bulk-spin autocorrelation function at $T=\infty $\ from
Gaussian decay to exponential decay for the anisotropy parameter $\Delta $
increasing from $0,$ and from exponential decay to power-law decay as $%
\Delta $ approaches 1, then replaced by a more rapid decay upon further
increase of $\Delta $ \cite{Bohm}. The dynamics of the equivalent-neighbor
XXZ model was studied in much detail at $T=0$ and $T=\infty $ using
different calculational techniques, and was found the same long-time
asymptotic behavior for the correlation function \cite{Lee,Lee2}.

More efforts have been concentrated on the dynamics of random quantum spin
systems in the past decades which can be used for describing more real
materials such as ferroelectric crystals \cite{yingyong1,yy1,yy2,yy3} and
spin glasses \cite{yingyong2}. Florencio and Barreto\textit{\ }have\textit{\ 
}studied the random transverse\ Ising model and obtained that the system
undergoes a crossover between a collective-mode behavior and a central-peak
one while the exchange couplings or external fields satisfy the bimodal
distribution \cite{J. Florencio}. Later, the dynamics of the four-body
transverse Ising model has been investigated for the cases of bond and field
randomness \cite{four},\ following which, a series of spin systems such as
the XY model with zero magnetic, the transverse Ising model with Gaussian
distribution and two-dimensional transverse Ising model etc.\ have also been
studied \cite{xy1,xy2,Z.-Q. Liu,Zhen-Bo Xu,chen}.\ A recent study shows that
the next-nearest-neighbor interaction has\ a strong influence on the
dynamics of the Ising\ system \cite{yuan}. For the random XXZ model, the
spin correlations have been investigated by using exact diagonalization \cite%
{antiferromagnetic}, the real space renormalization group method \cite{T
infinite,random1,random2} and a finite-chain study \cite{random3,Heinrich
Roder} etc. Infinite temperature spin-spin correlation function has been
found to display exponential localization in space indicating insulating
behavior for large enough random fields \cite{T infinite}. The transverse
correlation function at $T=0$ has been found to exhibit a power-law decay to
exponential decay depending on the exchange disorder \cite{Heinrich Roder}.

In this paper, we investigate the effects of the random fields on the time
evolution of the quantum XXZ model in the high-temperature limit.\ We find
that the system with the random fields that satisfy the bimodal
distribution\ undergoes a crossover between a central-peak behavior and a
collective-mode one with increasing field when the anisotropy parameter $%
\Delta $ is small (e.g., $\Delta =0.01$), but the collective-mode behavior
vanishes as $\Delta $ approaches 0.4, then when $\Delta $ increases to $1.0$%
, the central-peak behavior vanishes and the system just shows a disordered
behavior.

The arrangement of this paper is organized as follows. In Sec. \ref{come on}
we give a simple introduction to the 1-D quantum XXZ model and the
recurrence relations method. In Sec. \ref{A ZA} we discuss the results, and
Sec. \ref{jiayou} contains a summary.

\section{MODEL AND METHOD\label{come on}}

The Hamiltonian of the 1-D quantum spin-1/2 XXZ model with external fields
can be written as%
\begin{equation}
H=-\frac{J}{2}\sum_{i}\left[ \left( \sigma _{i}^{x}\sigma _{i+1}^{x}+\sigma
_{i}^{y}\sigma _{i+1}^{y}\right) +\Delta \sigma _{i}^{z}\sigma _{i+1}^{z}%
\right] -\frac{1}{2}\sum_{i}B_{i}\sigma _{i}^{z},  \label{1}
\end{equation}%
where $\sigma _{i}^{\alpha }$ $\left( \alpha =x,y,z\right) $ are Pauli spin
operators, $J$ and $\Delta $ are the exchange coupling and the anisotropy
parameter, respectively. $B_{i}$ denote the external fields, which may be
regarded as random variables. Clearly, this Hamiltonian can describe two
special cases: the Ising model for which $\Delta =\infty $ and the isotropy
XY model where $\Delta =0$.

The spin autocorrelation function plays an important part in the study of
the dynamics of quantum spin systems. It is defined as%
\begin{equation}
C\left( t\right) =\overline{\left\langle \sigma _{j}^{x}\left( t\right)
\sigma _{j}^{x}\left( 0\right) \right\rangle }  \label{2}
\end{equation}%
where $\overline{\langle \cdots \rangle }$ denotes an ensemble average
followed by an average over the random variable. The corresponding spectral
density which is the Fourier transform of $C\left( t\right) $ can be
expressed as 
\begin{equation}
\Phi \left( \omega \right) =\int_{-\infty }^{+\infty }e^{i\omega t}C\left(
t\right) dt,  \label{3a}
\end{equation}%
and for the mathematical\textit{\ }simplicity, $\Phi \left( \omega \right) $
is able to be obtained as%
\begin{equation}
\Phi \left( \omega \right) =\underset{\varepsilon \rightarrow 0}{\lim }\left[
\text{Re}\int_{0}^{\infty }dtC\left( t\right) e^{-zt}\right] ,  \label{3b}
\end{equation}%
where $z=\varepsilon +i\omega ,$ $\varepsilon >0.$

The recurrence relations method has been proved to be very powerful in the
calculation of dynamic correlation function \cite{chen,Lee fangfa,Zhen-Bo
Xu,Z.-Q. Liu,four,J. Florencio,F,yuan}. Next, we will give a brief
introduction to this method.

Considering a Hermitian operator $\sigma _{j}^{x}\left( t\right) $ as a
dynamical variable and then expanding it with an orthogonal set in a Hilbert
space,%
\begin{equation}
\sigma _{j}^{x}\left( t\right) =\sum_{\nu =0}^{\infty }a_{\nu }\left(
t\right) f_{\nu },  \label{4}
\end{equation}%
where $a_{\nu }\left( t\right) $ are the time-dependent coefficients. There
exists the following set of recurrence relation for the basis vectors $%
f_{\nu },$ 
\begin{equation}
f_{\nu +1}=iLf_{\nu }+\Delta _{\nu }f_{\nu -1},\text{ \ }\nu \geq 0,
\label{5a}
\end{equation}%
\begin{equation}
\Delta _{\nu }=\frac{\left( f_{\nu },f_{\nu }\right) }{\left( f_{\nu
-1},f_{\nu -1}\right) },\text{ \ }\nu \geq 1,  \label{5b}
\end{equation}%
where $L\equiv \left[ H,\text{ }\right] $ is the quantum Liouvillian
operator, $\left( f_{\nu },f_{\nu }\right) $ $=\overline{\left\langle f_{\nu
}f_{\nu }^{\dagger }\right\rangle },$ $f_{-1}\equiv 0$ and $\Delta
_{0}\equiv 1.$ The coefficients $a_{\nu }\left( t\right) $ in Eq. $\left( %
\ref{4}\right) $ satisfy the relation:%
\begin{equation}
\Delta _{\nu +1}a_{\nu +1}\left( t\right) =-\dot{a}_{\nu }\left( t\right)
+a_{\nu -1}\left( t\right) ,\text{ \ }\nu \geq 0,  \label{6}
\end{equation}%
where $\dot{a}_{\nu }\left( t\right) =\dfrac{da_{\nu }\left( z\right) }{dt},$
$a_{-1}\left( t\right) \equiv 0.$

The spin autocorrelation function $C\left( t\right) $ can be expressed as
the form of moment expansion%
\begin{equation}
C(t)=\sum_{k=0}^{\infty }\frac{\left( -1\right) ^{k}}{\left( 2k\right) !}\mu
^{2k}t^{2k},  \label{C}
\end{equation}%
with%
\begin{equation}
\mu ^{2k}=\frac{1}{Z}\overline{\text{Tr}\sigma _{j}^{x}\left[ H,\left[
H,\cdots \left[ H,\sigma _{j}^{x}\right] \cdots \right] \right] }\text{,}
\label{ju}
\end{equation}%
where $\mu ^{2k}$ is the $2k$th moment of $C\left( t\right) $. Supposing
that the first Q moments have been calculated by Eq. (\ref{ju}), we can
obtain $C\left( t\right) $ by constructing the pad\'{e} approximate. Because
of the mathematical complexity, a finite number of moments can be got, the
large times of $C\left( t\right) $ are divergent, thus we just can discuss
the short times of $C\left( t\right) $.

By taking the inner product for $\sigma _{j}^{x}\left( t\right) $\ and $%
\sigma _{j}^{x}\left( 0\right) $, from Eq. (\ref{4}),\ we can find that $%
a_{0}\left( t\right) $\ is the spin autocorrelation function%
\begin{equation*}
a_{0}\left( t\right) =\overline{\left\langle \sigma _{j}^{x}\left( t\right)
\sigma _{j}^{x}\left( 0\right) \right\rangle }=C\left( t\right) .
\end{equation*}%
Applying the Laplace transform $a_{\nu }(z)=\int_{0}^{\infty }e^{-zt}a_{\nu
}\left( t\right) dt$\ $\left( z=\varepsilon +i\omega ,\varepsilon >0\right)
\ $to Eq.$\left( \ref{6}\right) ,$ $a_{0}\left( z\right) $\ in the
continued-fraction representation can be obtained as 
\begin{equation}
a_{0}\left( z\right) =\frac{1}{z+\dfrac{\Delta _{1}}{z+\dfrac{\Delta _{2}}{%
z+\cdot \cdot \cdot }}}.  \label{7}
\end{equation}%
Because we can only get a finite number of recurrants, it is necessary to
terminate it with some schemes. Here, we use the Gaussian terminator \cite%
{Gaussian,F} Which can best serve our problem. With the help of the Pad\'{e}
approximate and the Gaussian terminator, we can obtain the spin
autocorrelation function $C\left( t\right) $ and the corresponding spectral
density $\Phi \left( \omega \right) $ of the system, respectively.

\section{RESULTS AND DISCUSSIONS\label{A ZA}}

In order to investigate the spin autocorrelation function $C\left( t\right) $
for $\sigma _{j}^{x}$, we choose the zeroth basis vector $f_{0}$ $=\sigma
_{j}^{x}$. With the recurrence relation Eq. $\left( \ref{5a}\right) $, the
remaining basis vectors can be obtained: 
\begin{equation*}
f_{1}=B_{j}\sigma _{j}^{y}+J\Delta \sigma _{j}^{y}\sigma _{j-1}^{z}-J\sigma
_{j-1}^{y}\sigma _{j}^{z}-J\sigma _{j+1}^{y}\sigma _{j}^{z}+J\Delta \sigma
_{j}^{y}\sigma _{j+1}^{z},
\end{equation*}%
\begin{eqnarray*}
f_{2} &=&\left( \Delta _{1}-B_{j}^{2}-2J^{2}-2J^{2}\Delta ^{2}\right) \sigma
_{j}^{x}+2J^{2}\Delta \sigma _{j-1}^{x}+2J^{2}\Delta \sigma
_{j+1}^{x}+J^{2}\Delta \sigma _{j-1}^{x}\sigma _{j-2}^{y}\sigma _{j}^{y} \\
&&-J^{2}\Delta \sigma _{j-2}^{x}\sigma _{j-1}^{y}\sigma _{j}^{y}+J^{2}\sigma
_{j+1}^{x}\sigma _{j-1}^{y}\sigma _{j}^{y}-2J^{2}\sigma _{j}^{x}\sigma
_{j-1}^{y}\sigma _{j+1}^{y}+J^{2}\sigma _{j-1}^{x}\sigma _{j}^{y}\sigma
_{j+1}^{y} \\
&&-J^{2}\Delta \sigma _{j+2}^{x}\sigma _{j}^{y}\sigma _{j+1}^{y}+J^{2}\Delta
\sigma _{j+1}^{x}\sigma _{j}^{y}\sigma _{j+2}^{y}-2B_{j}J\Delta \sigma
_{j}^{x}\sigma _{j-1}^{z}+\left( B_{j-1}+B_{j}\right) J\sigma
_{j-1}^{x}\sigma _{j}^{z} \\
&&+\left( B_{j-1}+B_{j}\right) J\sigma _{j+1}^{x}\sigma _{j}^{z}+J^{2}\Delta
\sigma _{j-1}^{x}\sigma _{j}^{z}\sigma _{j-2}^{z}-J^{2}\sigma
_{j-2}^{x}\sigma _{j-1}^{z}\sigma _{j}^{z}+J^{2}\Delta \sigma
_{j+1}^{x}\sigma _{j}^{z}\sigma _{j-1}^{z} \\
&&-2B_{j}J\Delta \sigma _{j}^{x}\sigma _{j+1}^{z}-2J^{2}\Delta ^{2}\sigma
_{j}^{x}\sigma _{j-1}^{z}\sigma _{j+1}^{z}+J^{2}\Delta \sigma
_{j-1}^{x}\sigma _{j}^{z}\sigma _{j+1}^{z}-J^{2}\sigma _{j+2}^{x}\sigma
_{j}^{z}\sigma _{j+1}^{z} \\
&&+J^{2}\Delta \sigma _{j+1}^{x}\sigma _{j}^{z}\sigma _{j+2}^{z},
\end{eqnarray*}%
etc. The first three norms of the basis vectors are obtained as follows:%
\begin{equation*}
\left( f_{0},f_{0}\right) =1,
\end{equation*}%
\begin{equation*}
\left( f_{1},f_{1}\right) =\overline{B_{j}^{2}}+2\overline{J^{2}}+2\overline{%
J^{2}\Delta ^{2}},
\end{equation*}%
\begin{eqnarray*}
\left( f_{2},f_{2}\right) &=&\overline{\Delta _{1}^{2}}-2\overline{\Delta
_{1}B_{j}^{2}}+\overline{B_{j}^{4}}-4\overline{\Delta _{1}J^{2}}+\overline{%
B_{j-1}^{2}J^{2}}+2\overline{B_{j-1}B_{j}J^{2}}+6\overline{B_{j}^{2}J^{2}}+2%
\overline{B_{j+1}B_{j}J^{2}} \\
&&+\overline{B_{j+1}^{2}J^{2}}+12\overline{J^{4}}-4\overline{\Delta
_{1}J^{2}\Delta ^{2}}+12\overline{B_{j}^{2}J^{2}\Delta ^{2}}+24\overline{%
J^{4}\Delta ^{2}}+8\overline{J^{4}\Delta ^{4}}.
\end{eqnarray*}%
Then the continued fraction coefficients can be got from Eq. (\ref{5b}).

Next, numerical results of the spin autocorrelation functions $C\left(
t\right) $ and the spectral densities $\Phi \left( \omega \right) $ are
given when the external fields satisfy three types of distributions: bimodal
distribution, Gaussian distribution and double-Gaussian distribution. With
special values of the anisotropy parameter $\Delta $, the effects of the
random external fields\ on the dynamics of the given system are investigated
as follows.

\subsection{Bimodal distribution}

We first consider the case that the external fields satisfy the bimodal
distribution. 
\begin{equation}
P\left( B_{i}\right) =p\delta \left( B_{i}-B_{1}\right) +\left( 1-p\right)
\delta \left( B_{i}-B_{2}\right) .  \label{8}
\end{equation}%
For simplicity and without loss of generality, we choose the exchange
coupling $J=1.0,$ which sets the energy scale, and the external fields $%
B_{1}=1.8$, $B_{2}=0.2$. For $\Delta $ $=0.01,$ $0.1,$ $0.4$ and $1.0$, the
results of the spin autocorrelation function $C\left( t\right) $ and the
spectral density $\Phi \left( \omega \right) $ are shown in Fig. 1 and Fig.
2, respectively. The continued-fraction coefficients are presented in the
insets.

When $\Delta $ is small (e.g., $\Delta =0.01,$ $0.1$)$,$ the spin
autocorrelation function $\left[ \text{see in Figs. 1(a1), (a2)}\right] $,
changes from a monotonically decreasing behavior to a damped oscillatory one
as $p$ increases. When $p=0$,\ the exchange coupling energy is higher than
the external field energy. The interaction among the spins is stronger than
that between the spin and the external field. The dynamics is dominated by
the exchange coupling energy and shows a central-peak behavior.\ When $%
p=0.25 $, the spin autocorrelation function has a slight fluctuation and the
fluctuation becomes acute with the increase of the external fields. When $%
p=1 $, the value of the external field is larger than that of the exchange
coupling. The system behaves as the precession of independent spins about
the field and the exchange coupling causes a damping. Hence, the system
presents a collective-mode behavior. Figs. 2(b1) and (b2) show that the peak
of $\Phi \left( \omega \right) $ moves from $\omega =0$ to $2$ as $p$
increases, which also reaches the conclusion that the system undergoes a
crossover from the central-peak behavior to the collective-mode one with
increasing field when $\Delta $ is small.

Figure 1(a3) shows that when $\Delta =0.4$, the dynamics of the system
changes from a central-peak regime to a disordered behavior which is
intervenient between a central-peak one and a collective-mode one as $p$
increases. The spin autocorrelation function for $p=0$ decays monotonically
to 0 and the spectral density is now peaked at $\omega =0.$ So, the system
is at the central-peak regime where the dynamics is mostly dominated by the
exchange coupling. By comparing the curve for $p=0$ in Figs. 1(a1) or (a2)
to the one in Fig. 1(a3), we find that $C\left( t\right) $ for $\Delta =0.4$
decays faster than that for $\Delta =0.01$\ or $\Delta =0.1.$ As p increases
to 1.0, the system shows not a collective-mode behavior but a disordered
one. The spectral density displayed in Fig. 2(b3) tends to have an expansion
at high frequency.

When $\Delta =1.0$ [see Figs. 1(a4) and 2(b4)], the system presents a
disordered behavior as the concentration of $B_{1}$ increases, i.e., in this
case, the dynamics of the system can not be characterized by either behavior
singly. Specially the case of $p=0$ is very similar to the most-disordered
case mentioned in Ref. \cite{J. Florencio}. By comparing the results when $%
\Delta $ $=0.01$ to that of the 1-D XY model, we find that they are very
similar, so the effect of the anisotropy parameter can be basically ignored,
in other words,\ the dynamics of the system is governed by the competition
between spin-spin interactions and the external fields. Comparing Figs.
1(a1), (a2),\ (a3) and (a4), it can be found that as the anisotropy
parameter $\Delta $ increases, the crossover from the central-peak behavior
to the collective-mode one vanishes. When $\Delta =1.0,$ the system becomes
the 1-D quantum Heisenberg system. The anisotropy parameter together with
the external field and the exchange coupling decide the dynamic behaviors of
the system. The competition between the spin-spin interactions and the
external fields becomes very fierce\ which drives the system to be
disordered.

\subsection{Gaussian distribution}

In the case, the external fields satisfy the Gaussian distribution,%
\begin{equation}
P\left( B_{i}\right) =\frac{1}{\sqrt{2\pi }\sigma _{B}}\exp \left[ -\left(
B_{i}-B\right) ^{2}/2\sigma _{B}^{2}\right] ,  \label{9}
\end{equation}%
where $B$ is the mean value of the external fields, $\sigma _{B}$ is the
standard deviation. Here, we take the exchange coupling $J$ equal to $1.0$
and $B$ equal to $0.0,$ $0.5,$ $1.0,$ $1.5,$ $2.0$. When the anisotropy
parameter $\Delta =0.1,$ $0.4$ and $1.0$, the results of the spin
autocorrelation function and the spectral density are displayed in Fig. 3
for $\sigma _{B}=0.3$ and Fig. 4 for $\sigma _{B}=3.0.$

Figure 3(a1) shows that when the standard deviation is small $\left( \sigma
_{B}=0.3\right) $ and $\Delta =0.1$, the system shows two types of dynamics
as $B$ increases: the central-peak behavior and the collective-mode
behavior. In this case, the effect of $\Delta $ can be ignored basically,
the dynamics of the system changes according to the concentration of $B.$ As 
$\Delta $ increases from $0.1$ to $1.0$, a disordered behavior replaces the
central-peak behavior or the collective-mode behavior. Figs. 3(a2), (b2) and
(c2) also show that the effects of the external fields are to urge the
system to show a crossover when $\Delta =0.1$ and as $\Delta $ increases to
1.0, the system displays a disordered behavior.

When the standard deviation is large enough $\left( \sigma _{B}=3.0\right) $%
[see Fig. 4], the system just shows a disordered behavior which is something
in between the central-peak behavior and the collective-mode one regardless
of the anisotropy parameter. From Fig. 3 and Fig. 4, it is not difficult to
see that both the crossover and disordered behavior are replaced by one
disordered behavior with the increase of $\sigma _{B}.$\ This is because the
value of the external field is large and the value range is wide\ when $%
\sigma _{B}=3.0$. The large external fields drive the spin orientation of
the system to be disordered.

\subsection{Double-Gaussian distribution}

Double-Gaussian distribution is a common form of bimodal distribution and
Gaussian distribution, which can be used to describe both discrete
distribution and continuous one,%
\begin{equation}
P\left( B_{i}\right) =p\frac{1}{\sqrt{2\pi }\sigma _{B}}\exp \left[ -\left(
B_{i}-B_{1}\right) ^{2}/2\sigma _{B}^{2}\right] +\left( 1-p\right) \frac{1}{%
\sqrt{2\pi }\sigma _{B}}\exp \left[ -\left( B_{i}-B_{2}\right) ^{2}/2\sigma
_{B}^{2}\right] ,  \label{10}
\end{equation}%
where $0\leq p\leq 1$ represents the concentration of $B_{1}$ that satisfies
the Gaussian distribution. The external fields satisfy Eq. $\left( \ref{10}%
\right) $, in which the mean values $B_{1}=1.8$ and $B_{2}=0.2,$ and the
exchange coupling is constant ($J=1.0$)$.$ $C\left( t\right) $ and $\Phi
\left( \omega \right) $ for $\Delta =0.1,$ $0.4$ and $1.0$\ are calculated
and the results are shown in Fig. 5 and Fig. 6, respectively. The figures
indicate that when $\sigma _{B}=0.3$, the system for $\Delta =0.1$ shows a
crossover between a central-peak behavior and a collective-mode one, and a
disordered behavior as $\Delta $ increases to $1.0$. However, the system
only shows a disordered behavior when $\sigma _{B}=3.0$, no matter what $%
\Delta $ takes.

From above discussion, we can see that the dynamical behavior of the system
is affected by the competition between the spin-spin interactions and the
external fields, not by the different disordered distribution. Also, it is
easy to find that the dynamics of the system is similar to that of the 1-D
quantum XY model \cite{Zhen-Bo Xu} when\ $\Delta $ is small (e.g., $\Delta
=0.01$). When $\Delta =0,$ the XXZ model becomes the isotropy XY model. We
find that the above results are the same as those in Ref. \cite{Zhen-Bo Xu}
when we take $\Delta =0.$

\section{SUMMARY\label{jiayou}}

We have studied the dynamics of the 1-D spin-$1/2$ quantum XXZ model in the
random external fields at the high-temperature limit by means of the
recurrence relations method. We find that the dynamics of the system with
three types of random distributions are affected by the competition among
the external field, the anisotropy parameter and the exchange coupling, but
the anisotropy parameter can be basically ignored when it is small (e.g., $%
\Delta =0.01$). For the case of bimodal disorder, when the anisotropy
parameter $\Delta $ is small (e.g., $\Delta =0.01$), the dynamics of the
system undergoes a crossover between a collective-mode behavior and a
central-peak one with increasing field; as $\Delta $ increases to 0.4, the
dynamics of the system changes from a central-peak regime to a disordered
behavior which is intervenient between a central-peak one and a
collective-mode one with the increase of the fields; then\ as $\Delta $
approaches 1, the system shows a disordered behavior. In the cases of
Gaussian disorder and double-Gaussian disorder, when the standard deviation
of the random field $\sigma _{B}$ is small, a\ disordered behavior replaces
the crossover as $\Delta $ increases. When $\sigma _{B}$ becomes large
enough, the system shows only a most disordered\ behavior regardless of the
anisotropy parameter$.$

\begin{acknowledgments}
This work was supported by the National Natural Science foundation of China
under Grant NO. 10775088, the Shandong Natural Science foundation under
Grant NO. Y2006A05, and the Science foundation of Qufu Normal University.
One of the authors (Yin-Yang Shen) thanks Shu-Xia Chen, Fu-Wu Ma, Hong Li
and Sha-Sha Li for useful discussions.
\end{acknowledgments}

\newpage

\begin{center}
\textbf{Figure Captions}
\end{center}

Fig.~$1$ The spin autocorrelation functions where the external fields take
the values $B_{1}=1.8$ with probability $p$ and $B_{2}=0.2$ with probability 
$\left( 1-p\right) .$ (a1), (a2), (a3), (a4) correspond to the cases of $%
\Delta =0.01,$ $0.1,$ $0.4$ and $1.0$. The continued-fraction coefficients
are presented in the insets.

\bigskip 

Fig.~$2$ The corresponding spectral densities for the same parameters as in
Fig. 1. (b1), (b2), (b3), (b4) correspond to the cases of $\Delta =0.01,$ $%
0.1,$ $0.4$ and $1.0$.

\bigskip 

Fig.~$3$ The spin autocorrelation functions and the spectral densities for
the case that the external fields satisfy the Gaussian distribution. The
standard deviation $\sigma _{B}$ takes $0.3.$ (a1), (a2) correspond to the
case that$\ \Delta $ equals to $0.1,$ (b1), (b2) correspond to the case that 
$\Delta $ equals to $0.4,$ and (c1), (c2) correspond to the case that$\
\Delta $ equals to $1.0.$

\bigskip 

Fig. $4$ The spin autocorrelation functions and the spectral densities for
the case that the external fields satisfy the Gaussian distribution. The
standard deviation $\sigma _{B}$ takes $3.0.$ (a1), (a2) correspond to the
case that$\ \Delta $ equals to $0.1,$ (b1), (b2) correspond to the case that 
$\Delta $ equals to $0.4,$ and (c1), (c2) correspond to the case that$\
\Delta $ equals to $1.0.$

\bigskip 

Fig. $5$ The spin autocorrelation functions for the case that the external
fields satisfy the double-Gaussian distribution in which the mean values $%
B_{1}=1.8$ and $B_{2}=0.2$ with probabilities $p$ and $\left( 1-p\right) $.
(a1), (a2) and (a3) correspond to the case that the standard deviation $%
\sigma _{B}$ takes $0.3$ and$\ \Delta $ takes the value $0.1$, $0.4,$ $1.0$,
respectively. (b1), (b2) and (b3) correspond to the case that$\ $the
standard deviation $\sigma _{B}$ takes $3.0$ and$\ \Delta $ takes the value $%
0.1$, $0.4,$ $1.0$, respectively.

\bigskip 

Fig. $6$ The spectral densities for the same parameters as in Fig. 5. (a1),
(a2) and (a3) correspond to the case that the standard deviation $\sigma _{B}
$ takes $0.3$ and$\ \Delta $ takes the value $0.1$, $0.4,$ $1.0$,
respectively. (b1), (b2) and (b3) correspond to the case that$\ $the
standard deviation $\sigma _{B}$ takes $3.0$ and$\ \Delta $ takes the value $%
0.1$, $0.4,$ $1.0$, respectively.

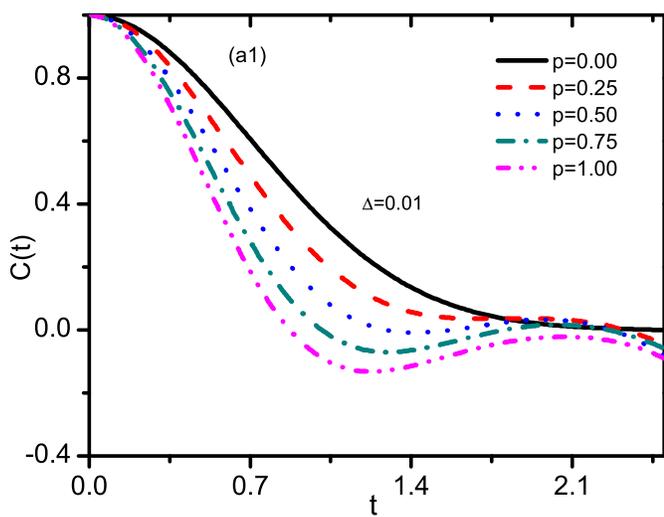
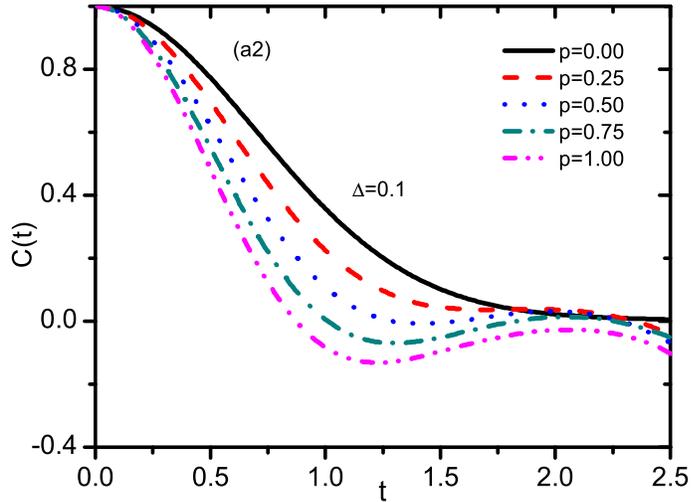
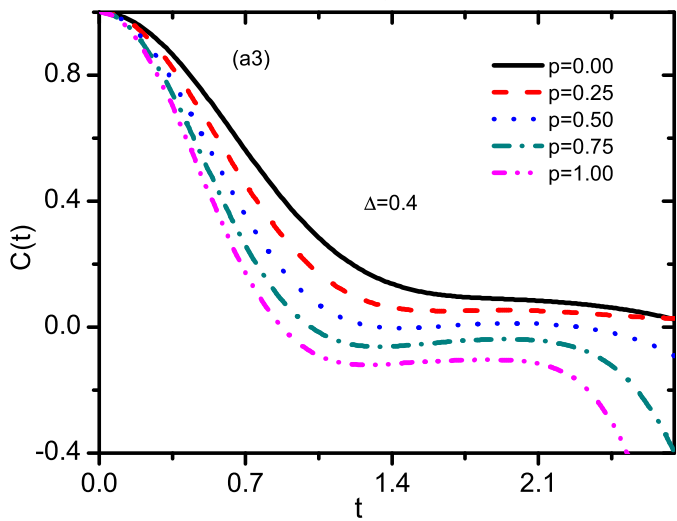
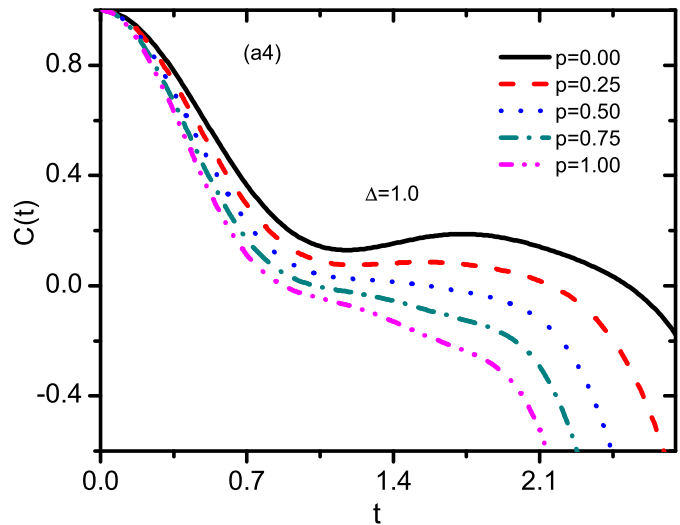
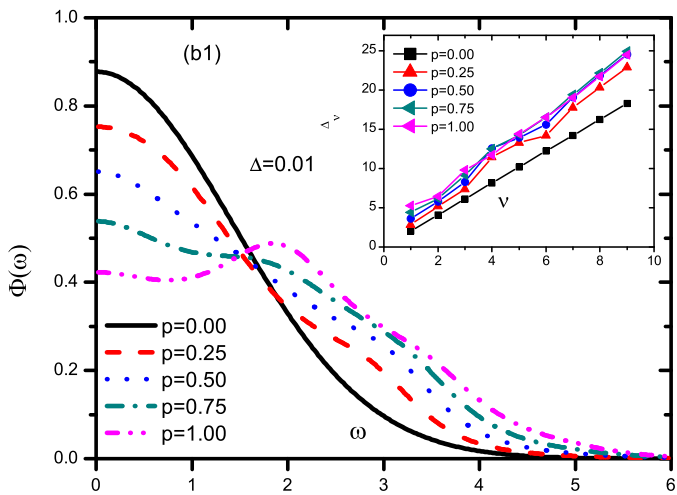
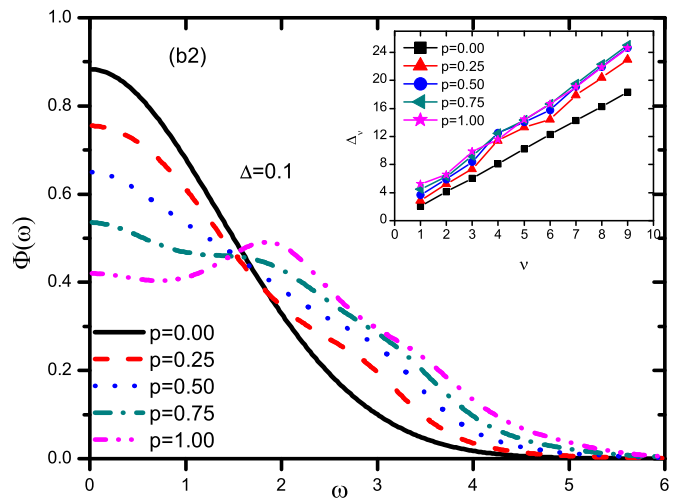
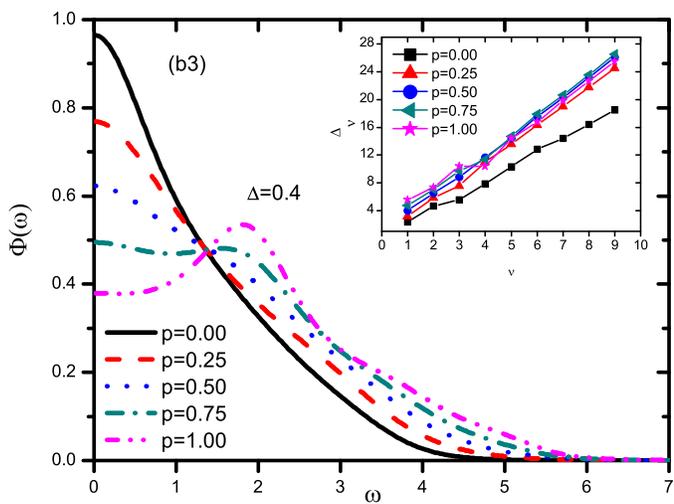
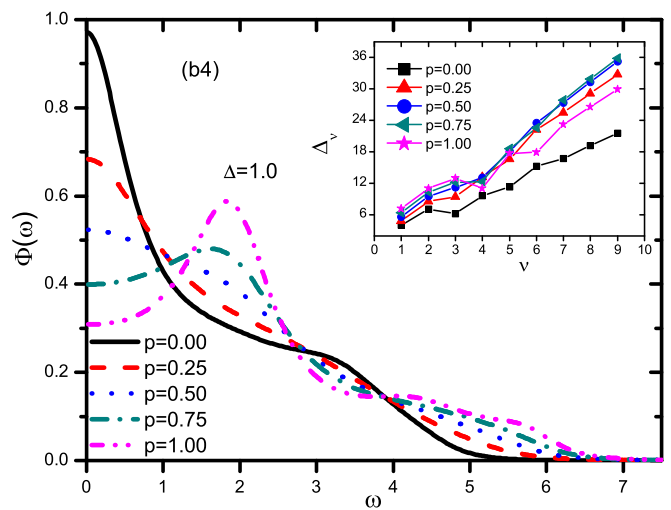
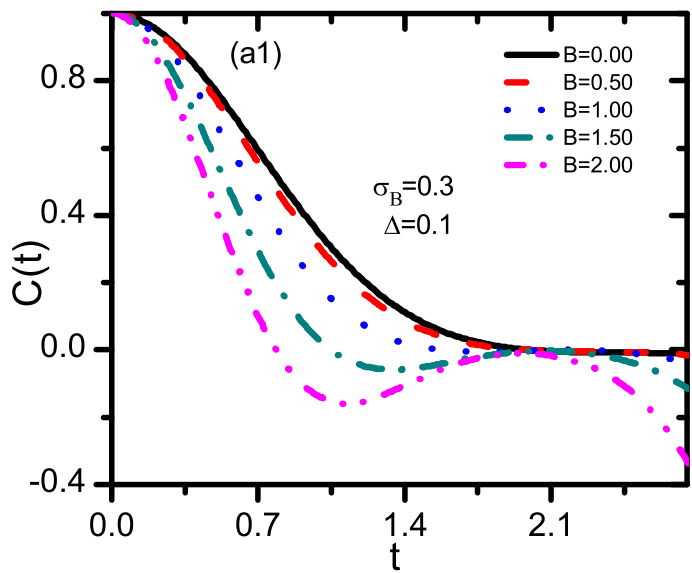
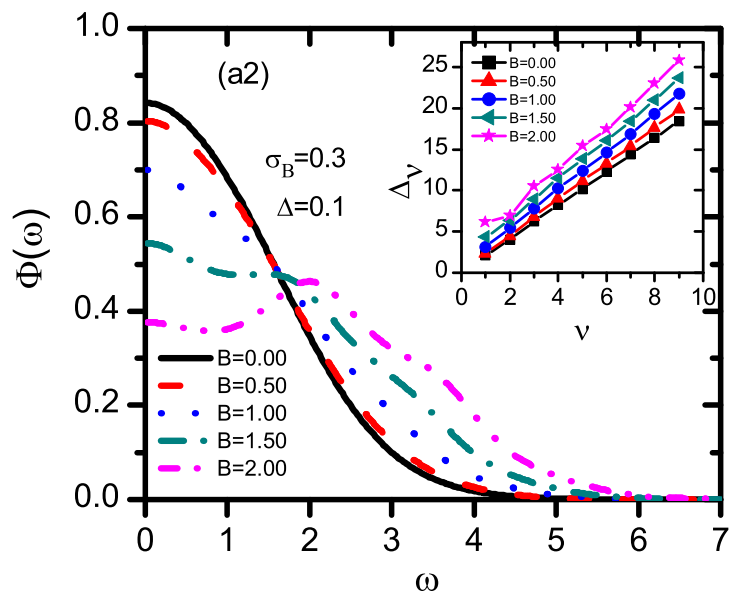
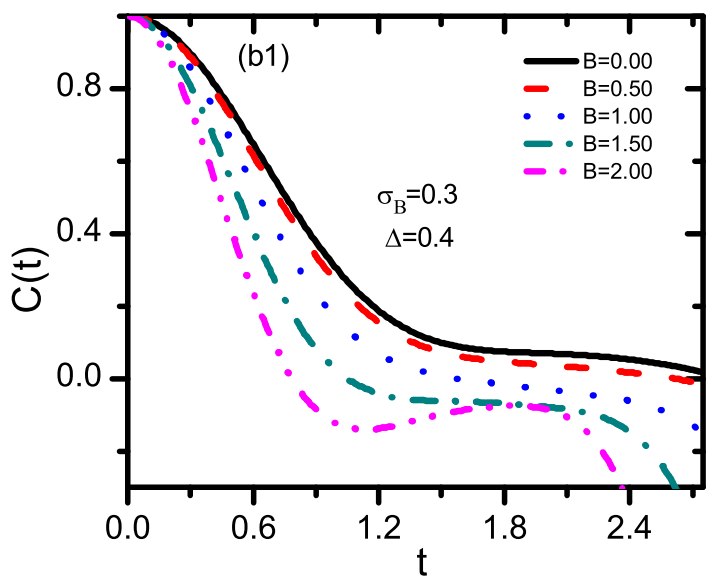
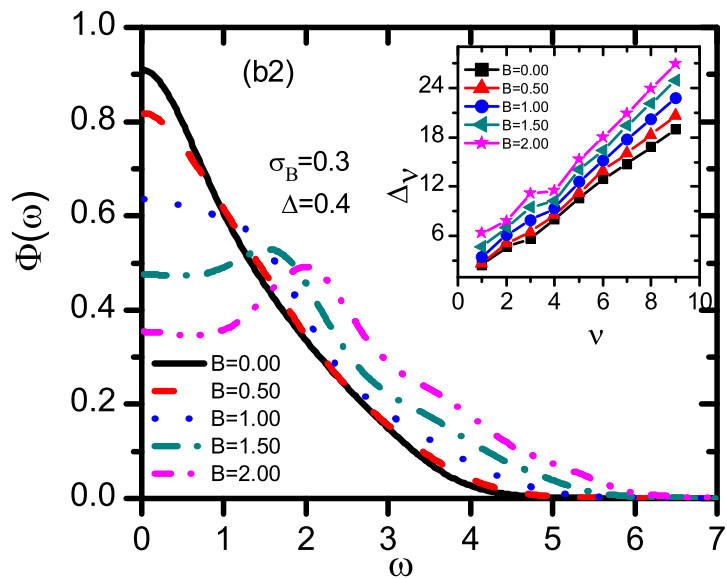
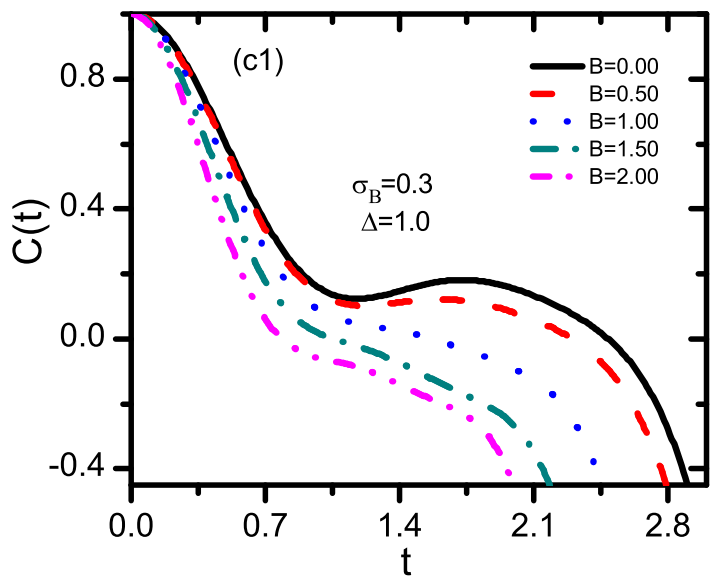
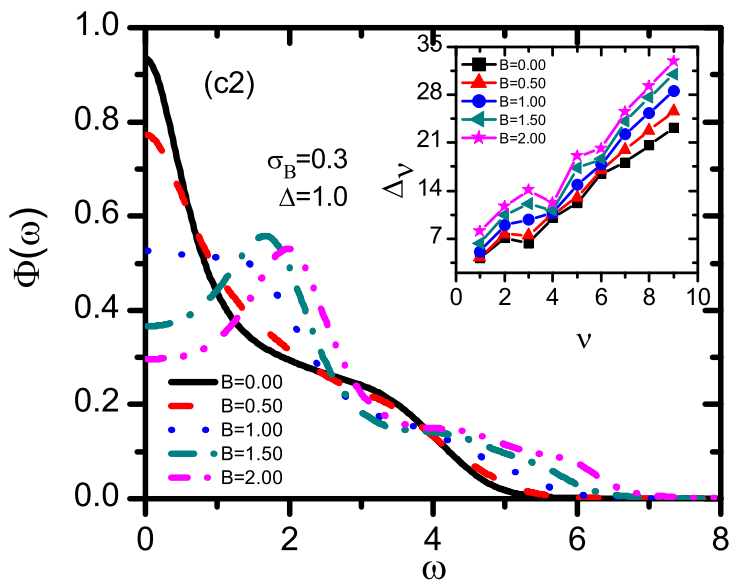
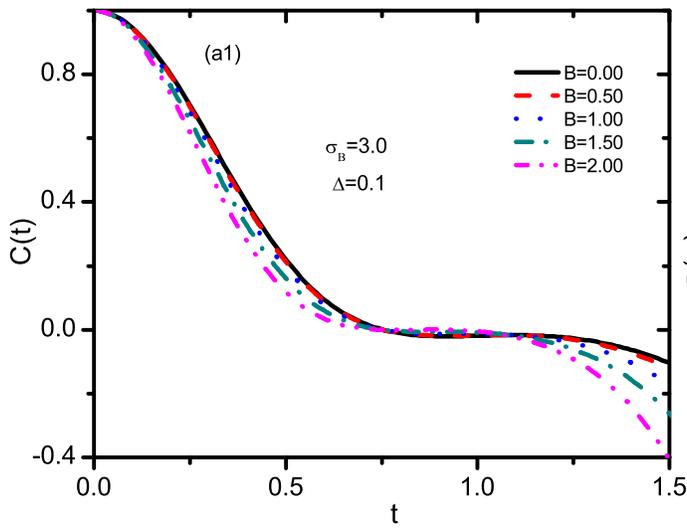
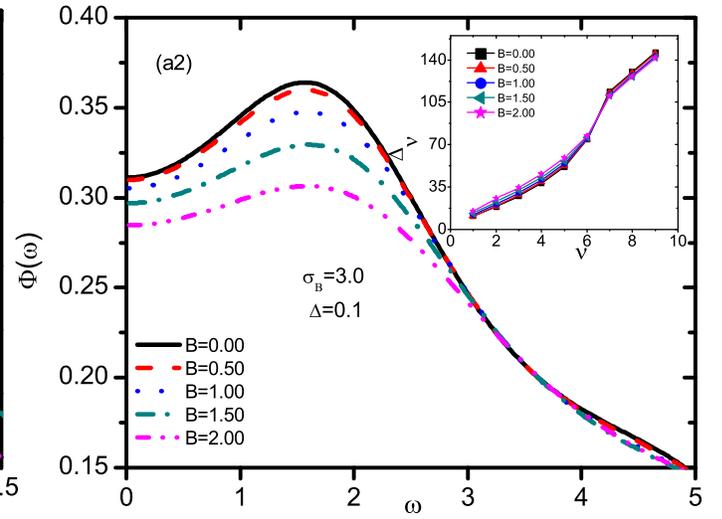
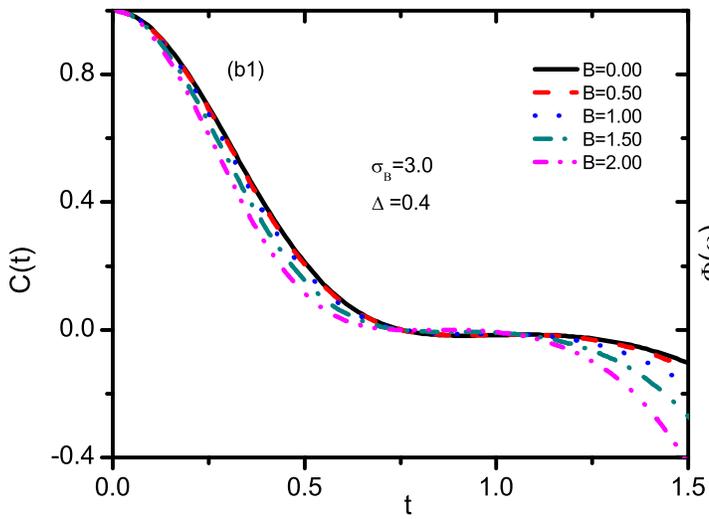
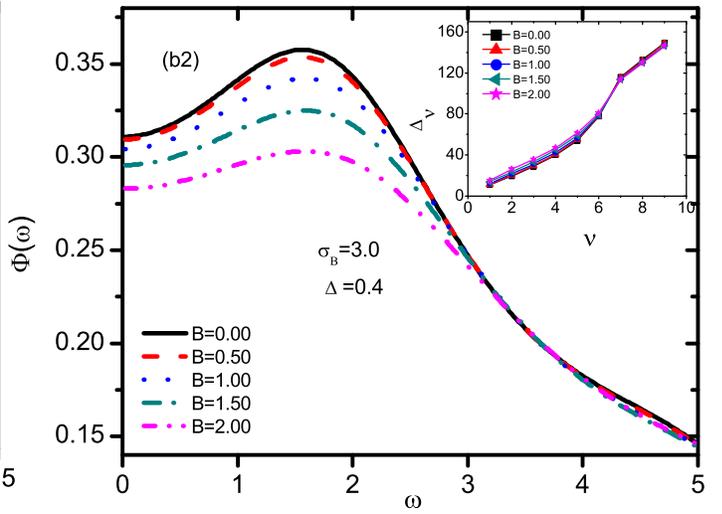
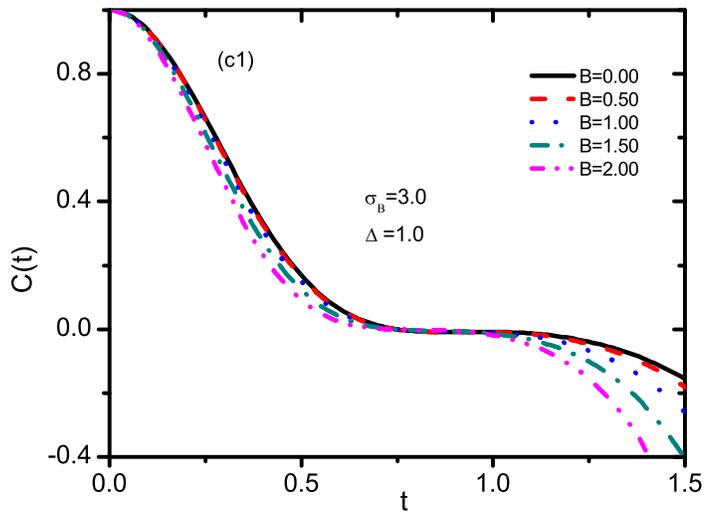
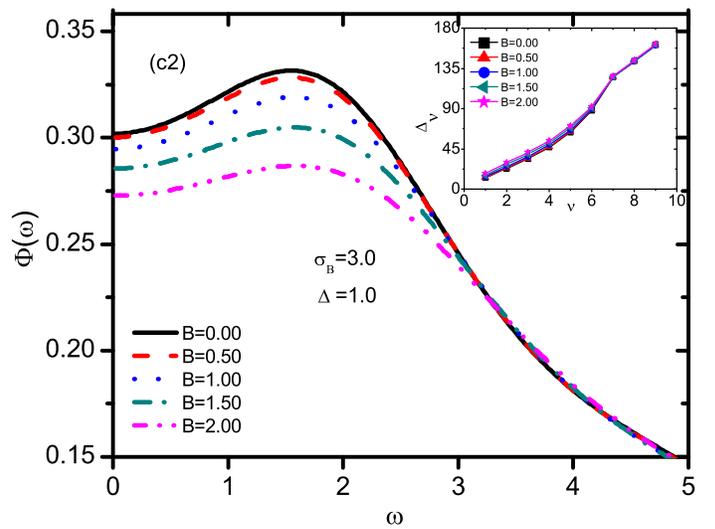
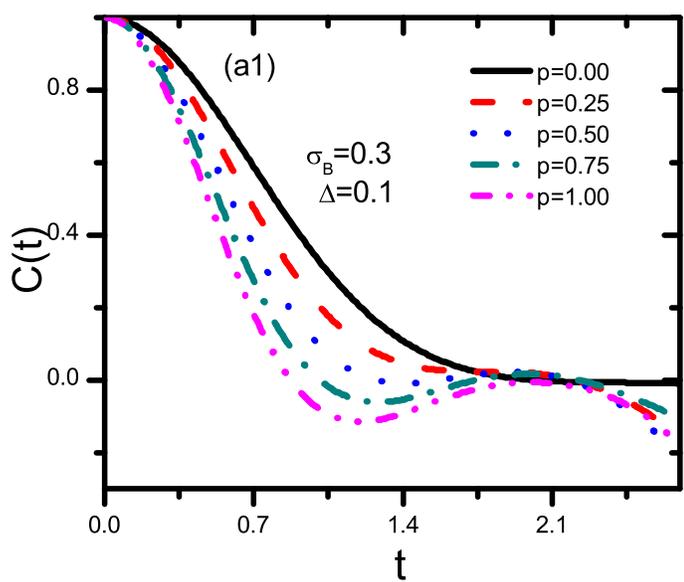
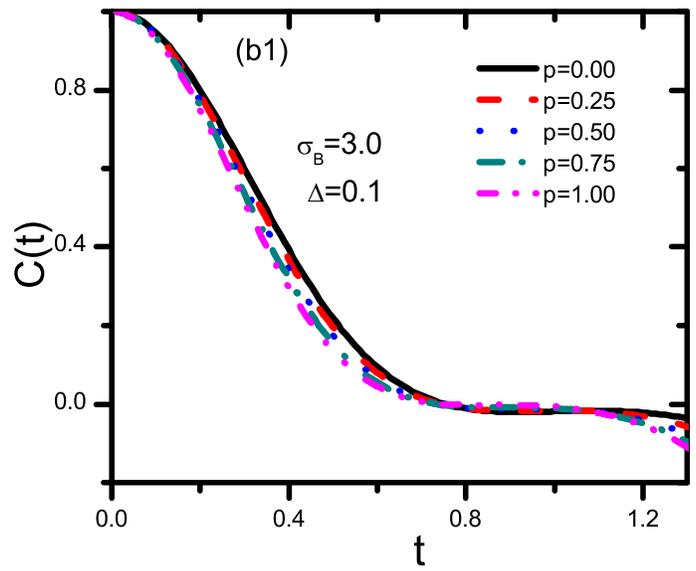
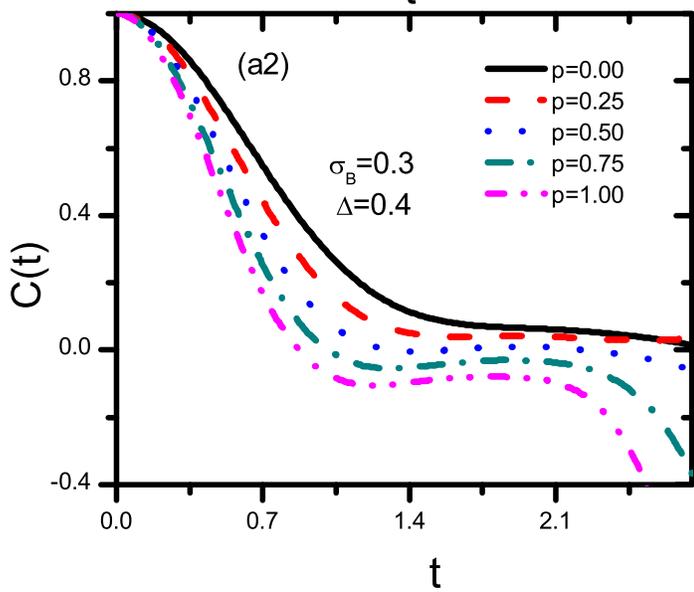
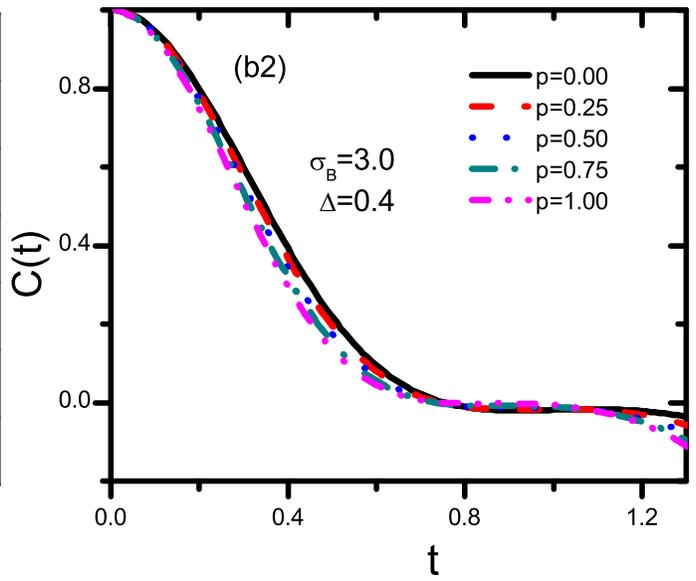
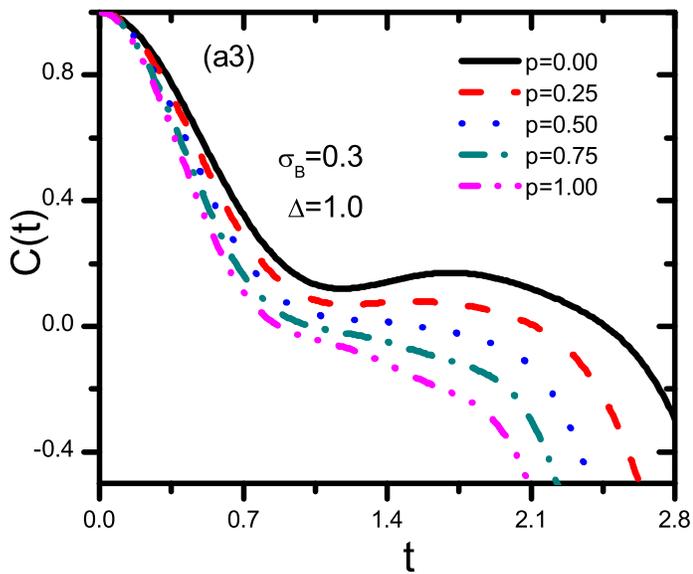
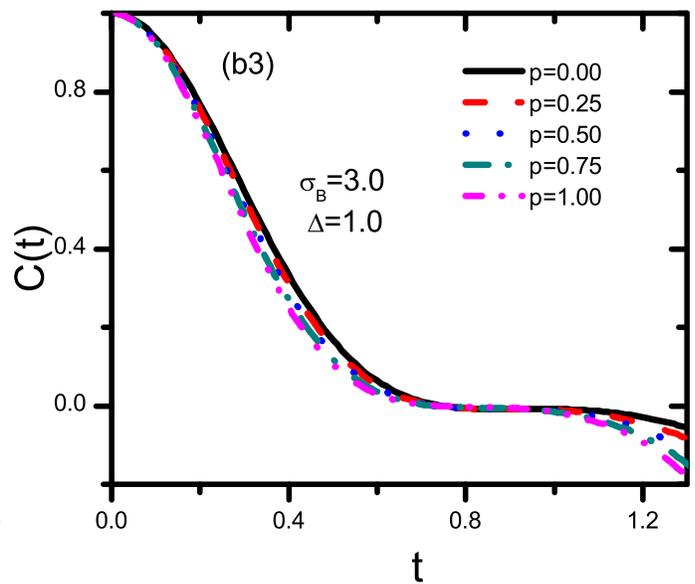
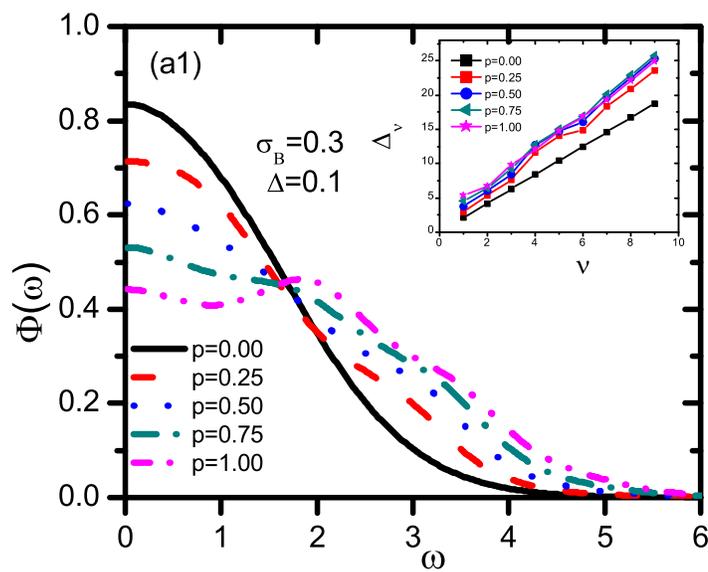
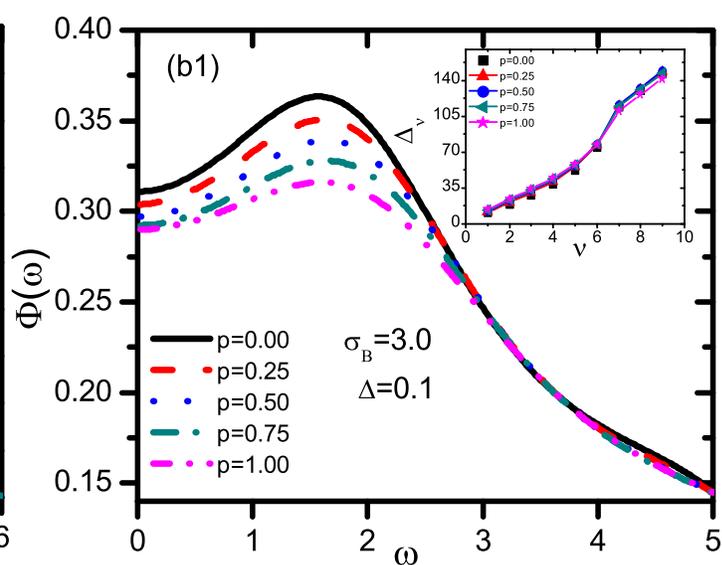
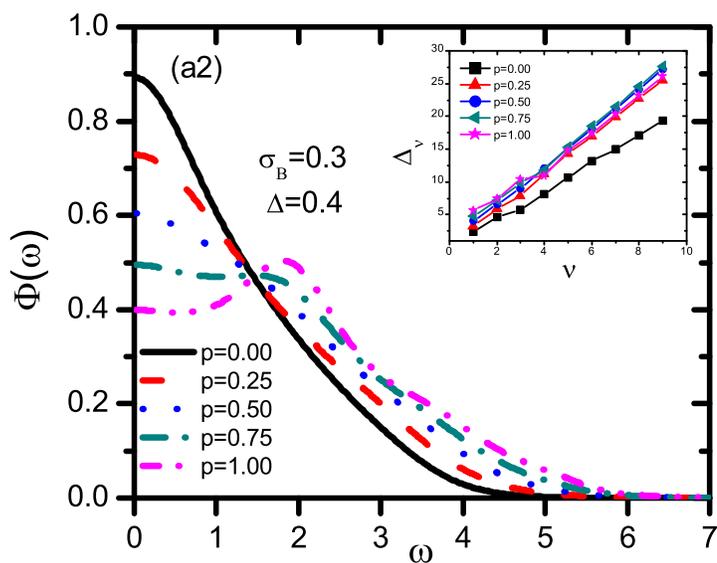
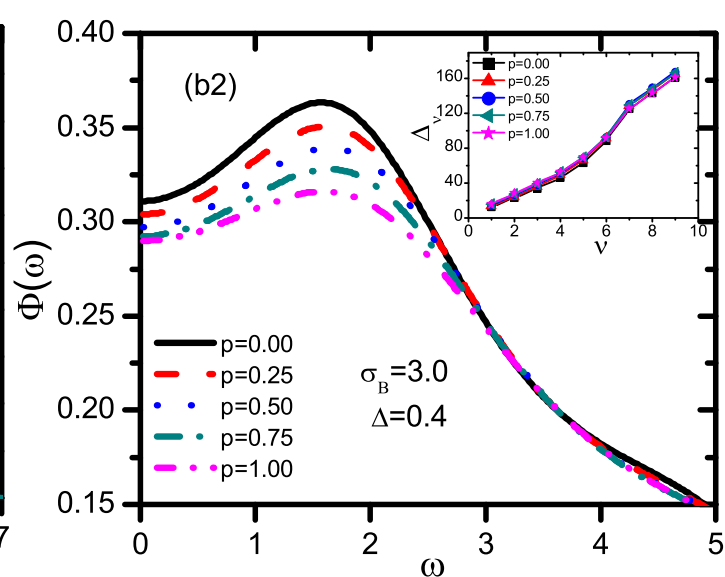
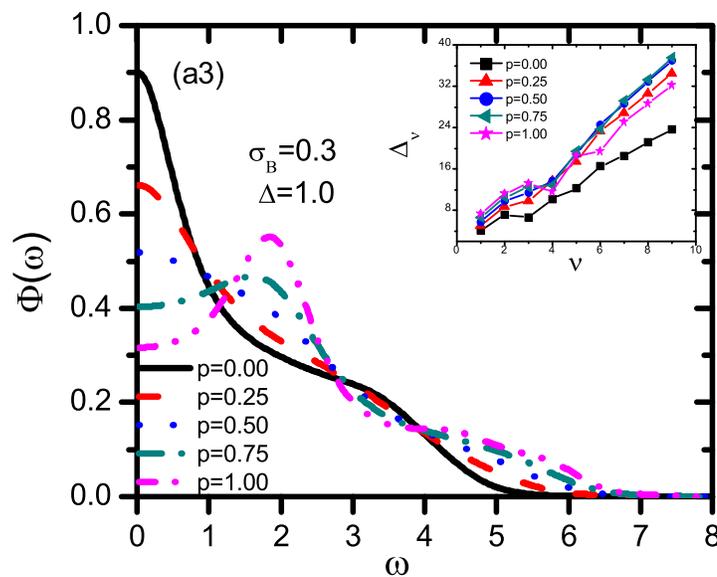
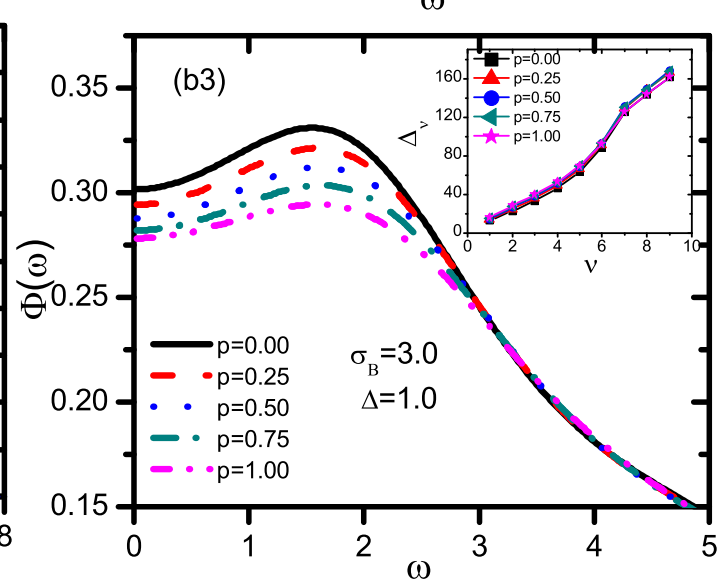
\end{document}